\documentclass[runningheads]{llncs}
\usepackage{times}
\usepackage{balance}
\usepackage{graphicx}
\usepackage{url}
\begin{document}
\title{Multi-Modality Image Inpainting using Generative Adversarial Networks}
%
%

\author{Aref Abedjooy\inst{1} \and
Mehran Ebrahimi\inst{2}\orcidID{0000-0002-3980-9582}}
\authorrunning{A. Abedjooy et al.}
%
\institute{Faculty of Science, Ontario Tech University,
Oshawa, Ontario, Canada \email{Aref.AbedjooyDivshali@ontariotechu.net} \and Faculty of Science, Ontario Tech University,
Oshawa, Ontario, Canada \email{Mehran.Ebrahimi@ontariotechu.ca}}
%
\maketitle             
%

\begin{abstract}
Deep learning techniques, especially Generative Adversarial Networks (GANs) have significantly improved image inpainting and image-to-image translation tasks over the past few years. To the best of our knowledge, the problem of combining the image inpainting task with the multi-modality image-to-image translation remains intact. 
In this paper, we propose a model to address this problem. The model will be evaluated on combined night-to-day image translation and inpainting, 
along with promising qualitative and quantitative results.
\keywords{Image-to-image translation \and Image inpainting \and Generative adversarial network \and Deep learning}
\end{abstract}

\section{Introduction}
Image-to-image translation and image inpainting are both challenging tasks.  Translating an image from one form, or modality, to another may involve generating an entirely new and realistic version of the image. 
Image-to-image translation has different application domains including digital arts, medical imaging \cite{kong2021breaking}.
The image inpainting task entails filling in missing regions in an image so that the whole image appears realistic. Image inpainting is an important step in many photo editing tasks. For example, an image would have a missing area after the removal of an unwanted object.

An interesting challenge would be to translate and inpaint images at the same time. In both tasks, the model is required to generate realistic outputs.
For example one may pose the problem of recovering an ideal day-time image of a scene, given a night-time image of the same scene where parts of the image is also missing. The combined image inpainting and translation tasks can be more challenging to create a realistic image. 

Deep Learning techniques have been successful at image-to-image translation and image inpainting tasks and we plan to combine and apply the existing generative models to address the problem as described int he following Sections.

In the following Sections we cover the related work, followed by the methodology, experiments and results, ablation study, conclusions and discussions. 


\section{Related work}

To the best of our knowledge, no studies have been conducted to combine image inpainting and image-to-image translation problems using Deep Learning (DL). Image generation with generative adversarial networks (GANs) has gained remarkable progress recently. A wide variety of image synthesis tasks, such as image editing, image composition, etc., have been investigated extensively using GANs. GANs provide plausible results for image inpainting specifically \cite{Zhao_2020_CVPR}\cite{zhou2020learning} \cite{li2020recurrent}\cite{yi2020contextual} \cite{Chen2021} \cite{nazeri2019edgeconnect}. Solving various image-to-image translation tasks with conditional GANs was introduced by \emph{pix2pix} \cite{isola2018imagetoimage}. Several studies have been conducted since then regarding the use of GANs to translate images \cite{park2020cut} \cite{CycleGAN2017} \cite{richardson2021encoding} \cite{baek2021rethinking}
\cite{emami2020spa}.

\section{Methodology}\label{methodology}
In this Section, we present the problem description along with a possible methodology to address the problem.
\begin{figure*}[!ht]
\includegraphics[width=\textwidth]{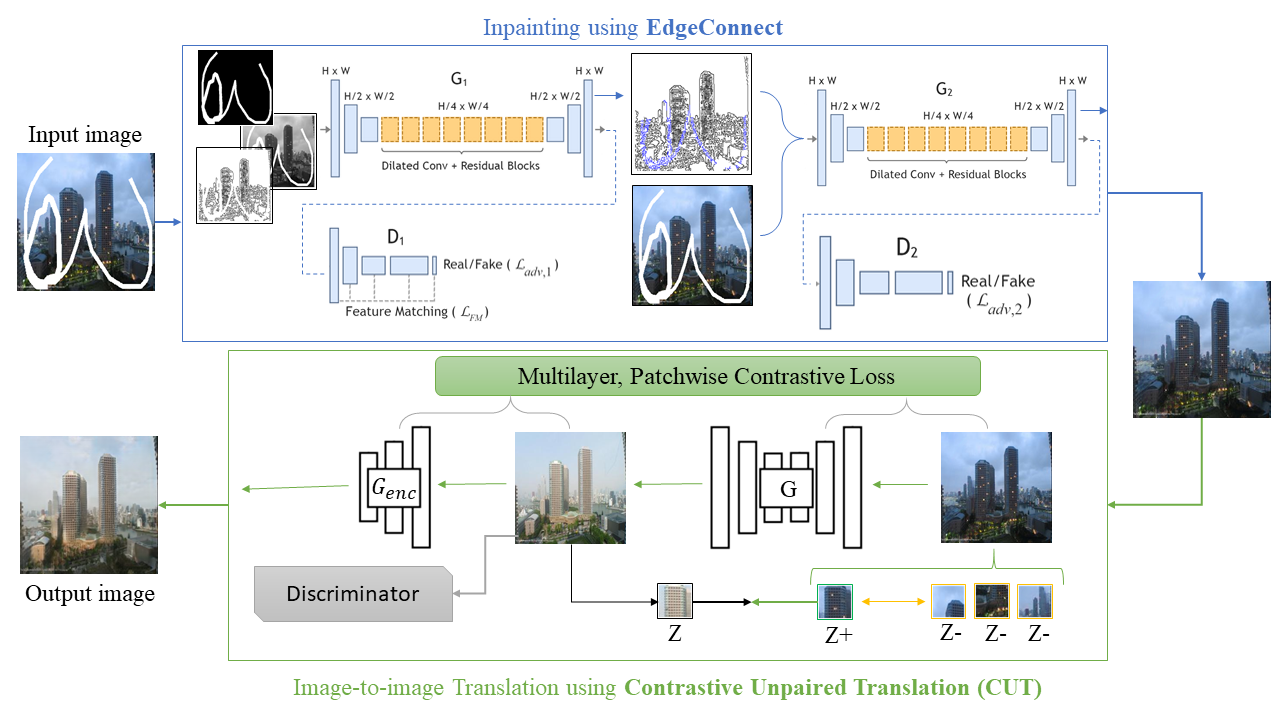}
\caption{Inpainting-First Model (M1), Inpainting module followed by an image-to-image translation module.} \label{fig:M1-A1}
\end{figure*}
\subsection{Problem description} \label{ProblemDescription}
The task of combining image inpainting and image-to-image translation can be formalized as follows. Here we introduce the problem in the context of day and nighttime images, that can be generalized to any arbitrary multi-modality image inpainting scenario. Given an incomplete input image in the nighttime mode $N$, 
with an arbitrary missing region 
, recover a corresponding complete daytime image $D$ in which the missing region $N_{m}$ has been completed with plausible content.

Human perception is perhaps the the most reliable tool for judging the perceived quality of a generated image. Ideally, the recovered image $D$ should meet the following criteria:  It should contain meaningful coherent content with $N$ so that human observer would accept it as a daytime image of $N$. It should have realistic-looking texture and color spectrum based on a daytime image. Finally, if the ground truth daytime image 
is available, the recovered image should ideally be similar to
the ground truth based on suitable image similarity measures. 
\subsection{System overview}
To address the described problem, we propose and investigate a two-stage model.
\subsubsection{Inpainting first model} The model involves first inpainting $N$ using some inpainting method, e.g., the \textbf{EdgeConnect} \cite{nazeri2019edgeconnect} to obtain a complete nighttime image $N_{EC}$ first. Once this is done, the new inpainted nighttime image without missing regions is translated into a realistic daytime image of the scene $D$ using an image translation approach, e.g., the \textbf{CUT} \cite{park2020cut} which is newer version of \emph{CycleGAN}  \cite{CycleGAN2017}. The inpainting-first model is illustrated in Figure \ref{fig:M1-A1}. In the following Sections, we will also investigate the effect of switching the order of the inpainting and translation modules in the recovery process, i.e., considering a translate-first model.

\subsubsection{Training}
There are two stages of training for the model.

In the \textbf{Inpainting-first} model, also referred to as M1, the \emph{first training stage} involves training of the EdgeConnect model to inpaint a given $N$ to generate $N_{EC}$. The EdgeConnect is trained in three stages: 1) training the edge completion model, 2) training the inpainting model and 3) training the joint model. To train the EdgeConnect model, incomplete grayscale image $N_{gray}$, edge map $N_{edge}$, and missing region mask $N_{m}$ of training 
data are the inputs to the training at the edge model stage. The output would be the full edge map $N_{edge}$  which contains hallucinated edges in $N_{m}$. During the training phase, the full edge map of nighttime ground truth $N_{gt_{edge}}$ computed using Canny edge detector is used to compare with $N_{edge}$ to optimize the model. To train the inpainting model, predicted edge map $N_{edge}$ and incomplete color image $N$ are passed to generate $N_{EC}$. For the first stage of the model in this study the joint model of EdgeConnect was used \cite{nazeri2019edgeconnect}. 

In the \emph{second stage} of the \textbf{Inpainting first} model, the inpainted nighttime image $N_{EC}$ is translated into the corresponding daytime image using the CUT model to yield $D_{CUT}$. For training the CUT model in the \emph{second stage}, $N_{EC}$ of all training data is computed as the source domain and $D_{gt}$ is considered as the target domain  \cite{park2020cut}.

\section{Experiments and Results}
The proposed model is evaluated quantitatively and qualitatively in this Section. Data-set and setup are explained here as well.

\subsection{Dataset} 
The \emph{Transient Attributes dataset} \cite{Laffont14} was used to create the \textbf{night2day} dataset. Each original sample is of size $256 \times 512$ containing two $256 \times 256$ images, i.e., one day image along with its corresponding night image. These images have been divided into $20,110$ night images and $20,110$ corresponding day images. The dataset was then split into training, validation, and test sets.

\subsection{Experiments}
\begin{figure}[!ht]
\centering
	\includegraphics[width=\linewidth]{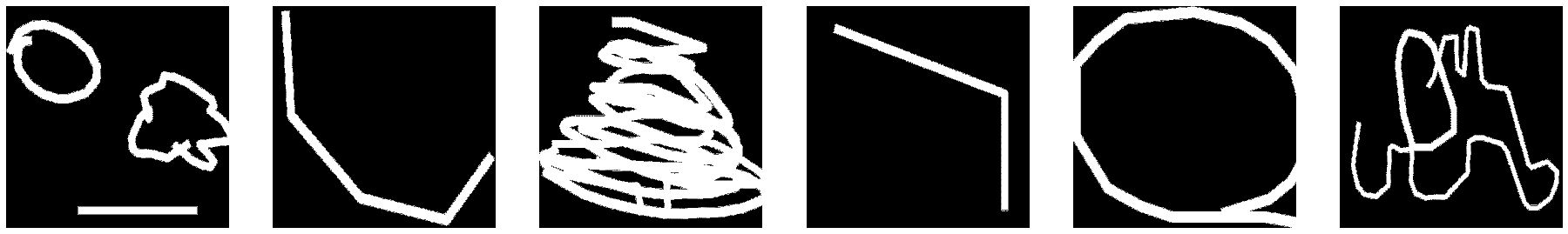}
	\includegraphics[width=\linewidth]{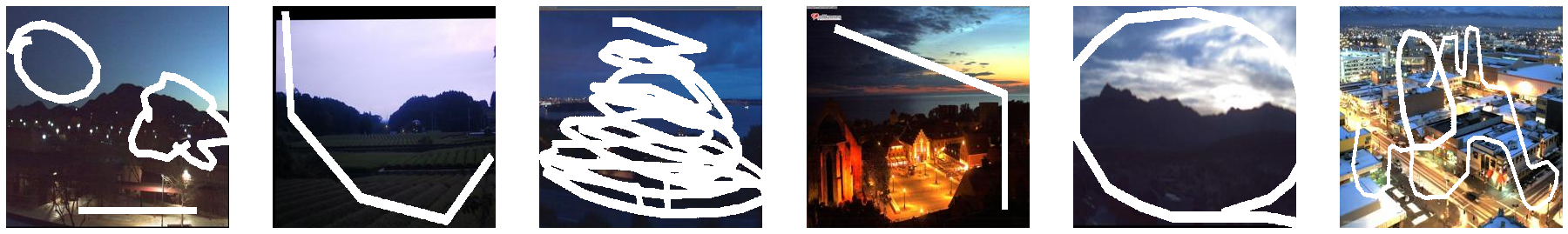}
	\caption{Examples of masks from the QD-IMD data \cite{iskakov2018semiparametric} applied to nighttime images.}
	\label{fig:QD-IMD}
\end{figure}
All experiments were performed on Linux Ubuntu operating system and an NVIDIA GeForce GTX TITAN X GPU.

\begin{figure}
\includegraphics[width=\textwidth]{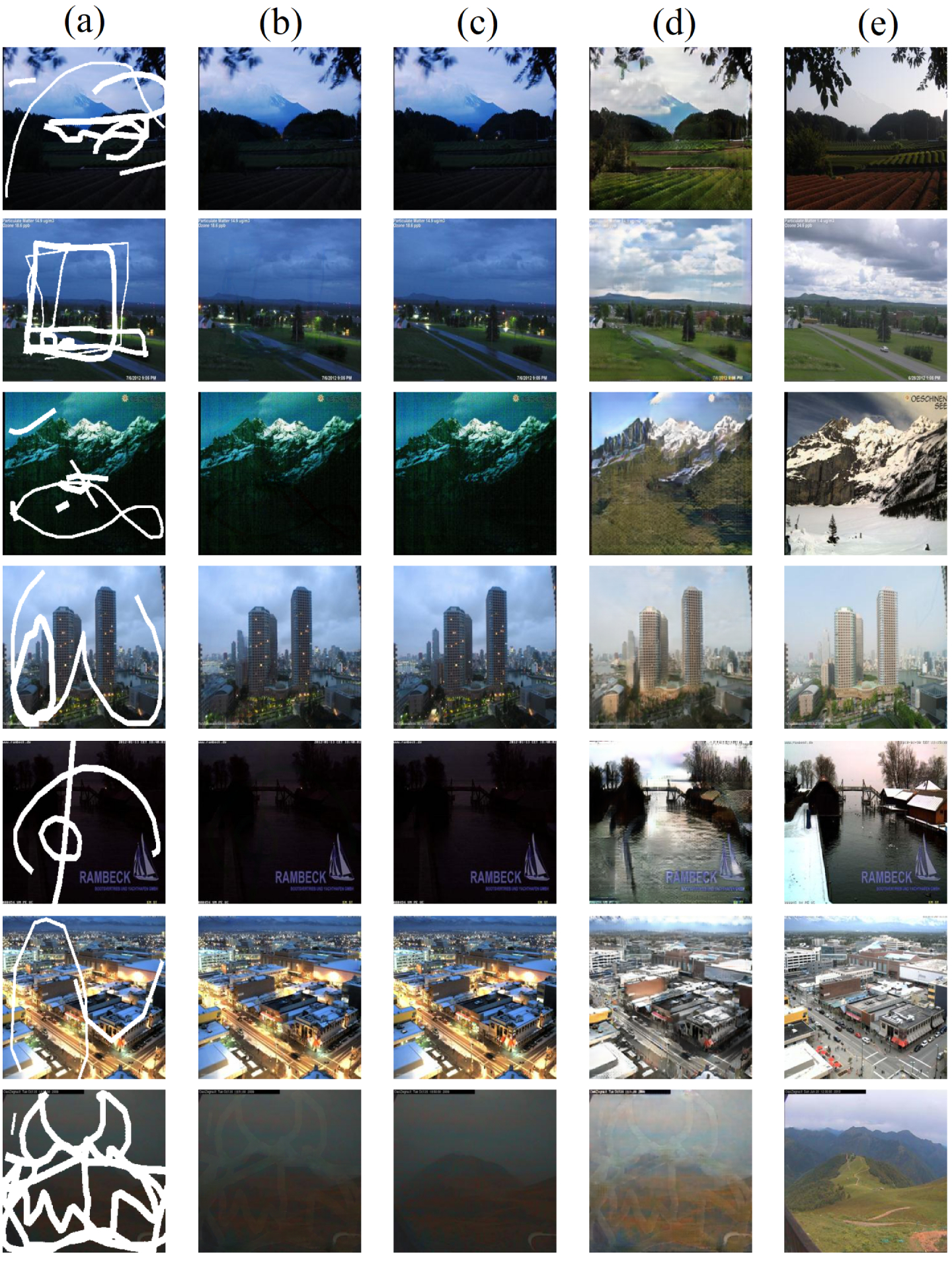}
\caption{Qualitative evaluation of model; from left to right; (a) Input image, (b) Inpainted night image, (c) Nighttime ground truth, (d) Generated output, (e) Daytime ground truth.}\label{fig:M1-A1:compare:qualitative}
\end{figure}

We used \textbf{Quick
Draw Irregular Mask Dataset (QD-IMD)} \cite{iskakov2018semiparametric} to apply random and irregular masks on night images. The QD-IMD research team believes that a combination of strokes drawn by the human hand is a good source of patterns for irregular masks \cite{iskakov2018semiparametric}. This dataset contains $60,000$ masks. The first row of Figure \ref{fig:QD-IMD} shows some examples of masks taken from the QD-IMD.

To produce nighttime images with missing points, these masks are randomly applied to night images. Therefore, $20,110$ night-time images with irregular missing points for training, validation, and test phases are generated. More precisely, $2,011$ or $10\%$ for testing, $2,011$ or $10\%$ for validation, and $16,088$ or $80\%$ for training phases. Second row of Figure \ref{fig:QD-IMD} illustrates examples of input nighttime images with missing points after applying the irregular masks. For each input image, the actual mask is provided to the model as well.

\subsection{Model Evaluation}\label{M1-A1_evl}

\begin{figure}
\includegraphics[width=\textwidth]{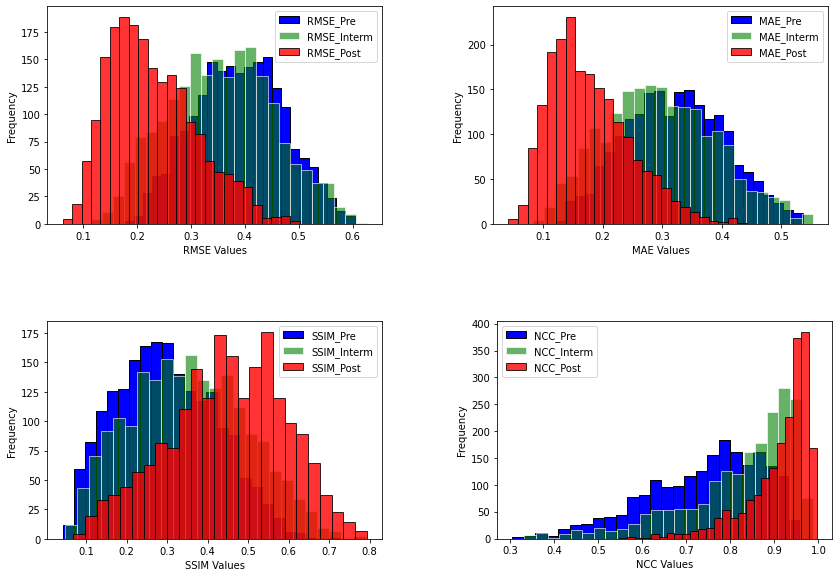}
\caption{Comparison of final results (Pre vs Intermediate vs Post) of the model using different measures.} \label{fig:M1-A1:compare:final_results}
\end{figure}

For testing the model, $2,011$ nighttime images with missing points are used. Since the images are of different modalities, i.e, night and day and we have provided a two-step approach, different evaluation phases have been defined for a more meaningful quantitative evaluation of the results as explained below;

\begin{itemize}
    \item \textbf{Phase 1}: Comparing \emph{the nighttime image with missing points (Input)} vs \emph{daytime ground truth image}, 
    \item \textbf{Phase 2}: Comparing \emph{the inpainted night image after inpainting} vs \emph{daytime ground truth image},
    \item \textbf{Phase 3}: Comparing \emph{the generated daytime image without missing points (output)} vs \emph{daytime ground truth image}.
\end{itemize}

\emph{Root Mean Square Error (RMSE)}, \emph{Mean Absolute Error (MAE)}, \emph{Structural Similarity Index Measure (SSIM)}, \emph{Normalized Cross-Correlation (NCC)}, and \emph{Fréchet inception distance (FID)} are calculated for each phase and histograms of these similarity measures over the test data are provided. 
\subsection{Results:}
The total performance of the model can be evaluated by comparing Phase 1 with Phase 2 and Phase 3.
We can refer to Phase 1 as \textbf{``Pre"} evaluation of similarities before applying the method. Results from Phase 2 are considered \textbf{``Intermediate"} after the first step, e.g, after inpainting. Finally phase 3 which compares the output with the ground truth, could be referred to as \textbf{``Post"}. For each measure, we compare the results of these phases. Figure \ref{fig:M1-A1:compare:final_results} illustrates these comparisons.

Figure \ref{fig:M1-A1:compare:final_results} shows that the \emph{RMSE} (top-left) and the \emph{MAE} (top-right) are shifted to the left  from the \emph{``Pre"} state to the \emph{``Post"} state. Lower value of these measures for final results in \emph{Phase 3} indicates better performance. Similarly, the \emph{Phase 3} or the \emph{``Post"} state provides higher values for \emph{SSIM} (bottom-left), and the \emph{NCC} (bottom-right) in comparison with the \emph{``Intermediate"} and the \emph{``Pre"} states.
Table \ref{table:M1-A1:compare:final_results} presents the \emph{mean} and the \emph{standard deviation} of these measures for \textbf{Pre}, \textbf{Intermediate}, and \textbf{Post}. The FID metric for each phase is computed as well. Based on all these measures, table \ref{table:M1-A1:compare:final_results} suggests that final results from the model are more consistent with ground truth.

\begin{table}
\begin{center}
\caption{Comparing final results (\textbf{Phase 1-A vs Phase 2-A vs Phase 3}) of the model using different measures.}\label{table:M1-A1:compare:final_results}
\begin{tabular}{  c | c | c | c }
& \textbf{Pre} & \textbf{Intermediate} & \textbf{Post} \\
\hline
\emph{RMSE}($\downarrow$)& 0.39$\pm$0.08&
0.36$\pm$0.10& \textbf{0.23}$\pm$\textbf{0.08}\\
\hline
\emph{MAE}($\downarrow$)& 0.16$\pm$0.06&
0.14$\pm$0.07& \textbf{0.06}$\pm$\textbf{0.04} \\
\hline
\emph{SSIM}($\uparrow$)& 0.30$\pm$0.12 & 
0.35$\pm$0.14 & \textbf{0.45} $\pm$\textbf{0.14} \\
\hline
\emph{NCC}($\uparrow$)& 0.74$\pm$0.13 & 
0.82$\pm$0.13 & \textbf{0.91}$\pm$\textbf{0.07} \\
\hline
\emph{FID}($\downarrow$)& 271.01 & 80.28 & \textbf{56.77}  \\ 
\hline
\end{tabular}
\end{center}
\end{table}
In order to \textbf{evaluate the results of the model qualitatively}, Figure \ref{fig:M1-A1:compare:qualitative} provides some sample test images at different stages. From left to right, each row contains a) Input i.e. the nighttime image with missing points,
b) The inpainted nighttime image without missing points (after inpainting),
c) Nighttime ground truth image,
d) Output i.e. the generated daytime image without missing points, and
e) Daytime ground truth image. 
It can be observed that when enough details exist in the input image, the model produces visually plausible results.

\subsection{Evaluation of a switched recovery order model }\label{M1-A2_evl}
\begin{figure}
\includegraphics[width=\textwidth]{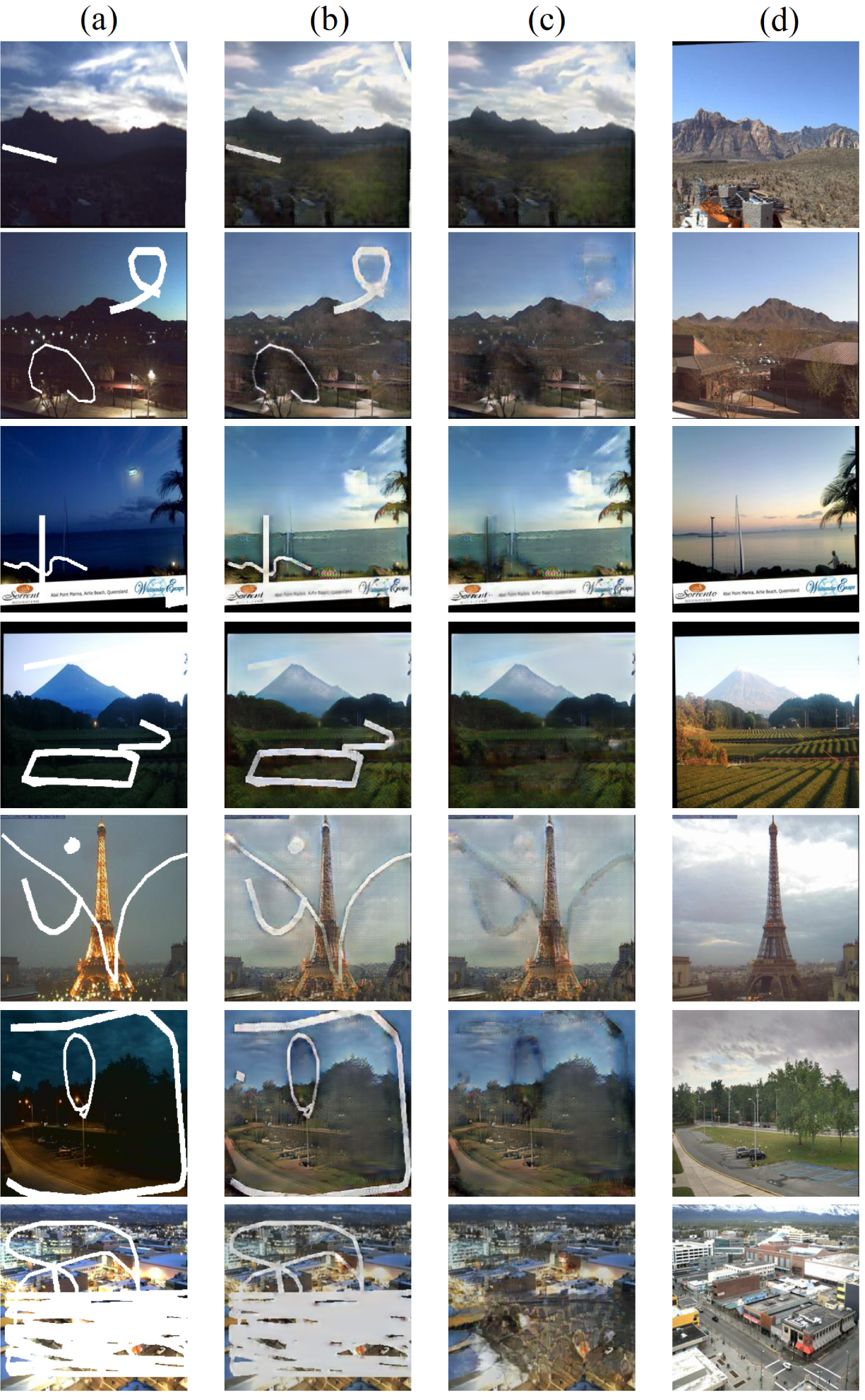}
\caption{Qualitative evaluation of the translation first  model; from left to right; (a) Input image, (b) Generated day image with missing points, (c) Output image, (d) Day ground truth.} \label{fig:M1-A2:compare:qualitative}
\end{figure}

A question may be posed whether the model produces similar results if we change the order of the inpainting and translation modules, i.e., if the translation module was applied first followed by the inpainting. In order to answer this question, the so called \textbf{M2: Translation first model} was implemented and its results are presented in Figure
\ref{fig:M1-A2:compare:qualitative}.

\begin{figure}
\includegraphics[width=\textwidth]{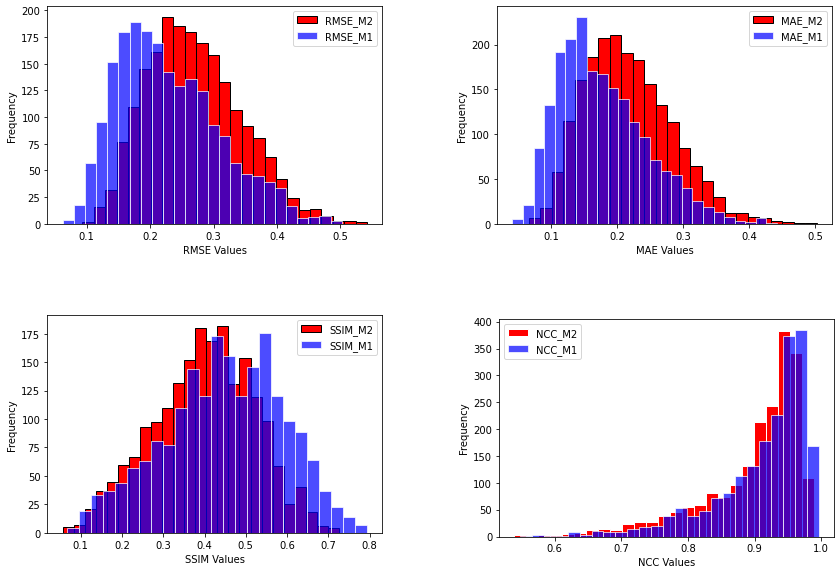}
\caption{Comparison of final results of the \emph{M1: Inpainting first} with the \emph{M2: Translation first} models using different measures.} \label{fig:M1-A1_VS_M1-A2:compare:final_results}
\end{figure}

The obtained results of \textbf{Phase 3} for the \emph{inpainting first} and the \emph{translation first} models allow us to compare their performances. In both models, phase 3 is comparing \emph{the generated daytime image (Output)} vs \emph{daytime ground truth image}. This comparison using the \emph{RMSE} (top-left), the \emph{MAE} (top-right), the \emph{SSIM} (bottom-left), and the \emph{NCC} (bottom-right) is illustrated in Figure \ref{fig:M1-A1_VS_M1-A2:compare:final_results}. The RMSE histogram and MAE histogram for the \emph{inpainting first model} are shifted to the \emph{left}, indicating superior results. Similarly, the SSIM, and NCC histograms are shifted towards the \emph{right}, which reaffirms better results based on these measures for the \emph{inpainting first model}. Table \ref{table:M1-A1_VS_M1-A2:compare:final_results} summarizes these comparisons.

\begin{table}
\begin{center}
\caption{Comparing final results of the \textbf{M1: Inpainting first model} with \textbf{M2: Translation first model} using different measures.}\label{table:M1-A1_VS_M1-A2:compare:final_results}
\begin{tabular}{  c || c | c  }
  &\textbf{M2}& \textbf{M1}\\
 \hline
   \emph{RMSE}($\downarrow$)& $0.27\pm0.07$& \textbf{0.23}$\pm$\textbf{0.08}\\
   \hline
   \emph{MAE}($\downarrow$)& $0.08\pm0.04$& \textbf{0.06}$\pm$\textbf{0.04}\\
   \hline
   \emph{SSIM}($\uparrow$)& $0.40\pm0.12$& \textbf{0.45}$\pm$\textbf{0.14}\\
   \hline
   \emph{NCC}($\uparrow$)& $0.90\pm0.07$& \textbf{0.91}$\pm$\textbf{0.07}\\
   \hline
   \emph{FID}($\downarrow$)& $108.62$&\textbf{56.77}\\ 
  \hline
\end{tabular}
\end{center}
\end{table}
A qualitative evaluation of the images generated using these two models, which are illustrated in Figure \ref{fig:M1-A1:compare:qualitative} and Figure \ref{fig:M1-A2:compare:qualitative}  as well as quantitative evaluations above confirm that the \textbf{inpainting first model} provides a better solution.

\section{Ablation Study}\label{ablation_study}

Mask line thickness increases the number of missing pixels as well as the distance between those pixels and known neighboring pixels along the edge of the mask. Those neighboring pixels provide crucial information for hallucinating the missing points and translating a complete night image into a coherent day image. 

\begin{figure}
\includegraphics[width=\textwidth]{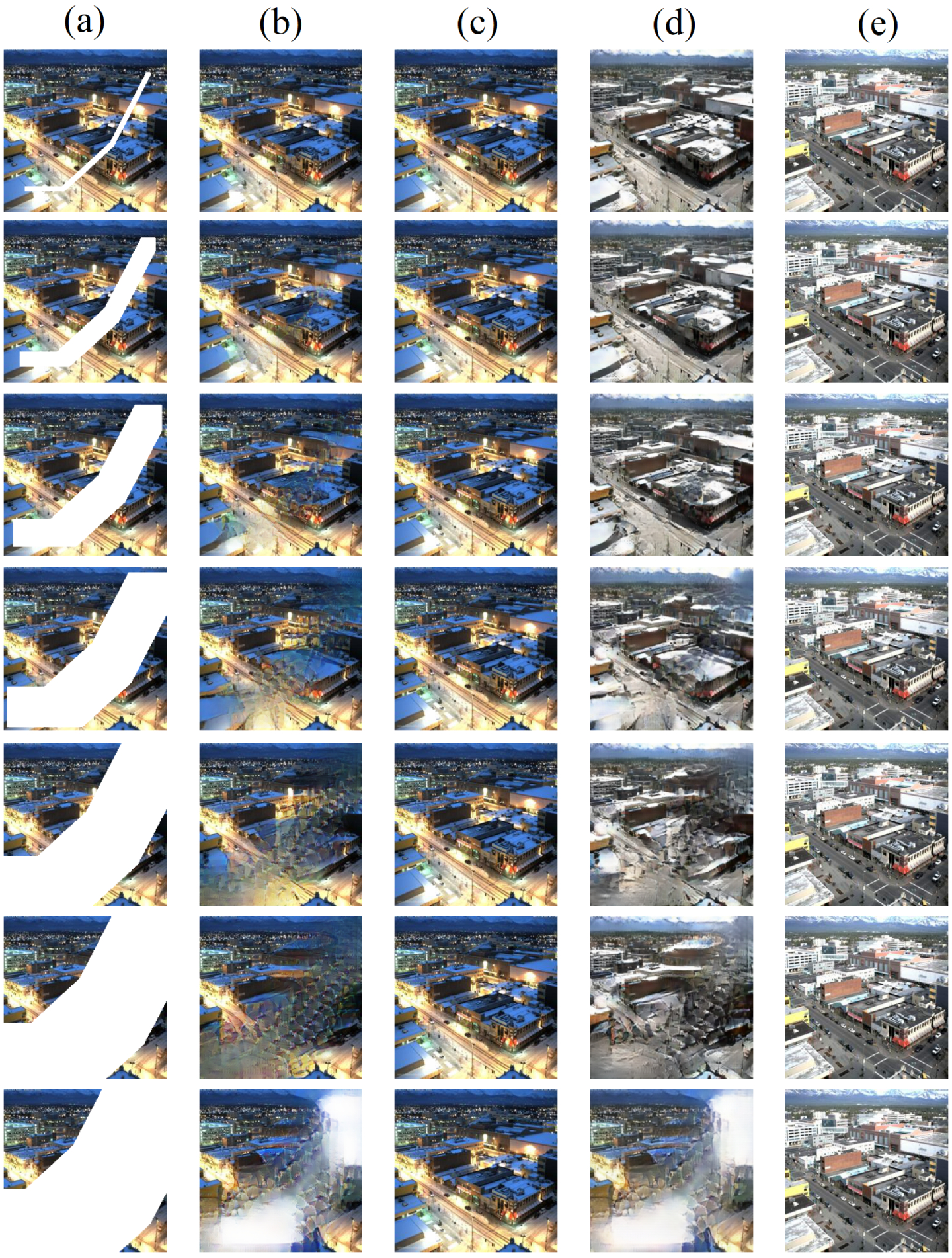}
\caption{Ablation study of the model; Analyzing mask dilation in input. From left to right; (a) Input image, (b) generated inpainted night image, (c) Night ground truth, (d) Output image, (e) Day ground truth.} \label{fig:ablation study}
\end{figure}

In order to study the effects of the mask on the model's performance, we choose an arbitrary night image from test data and apply a mask with a simple thin line from the \emph{QD-IMD dataset}  \cite{iskakov2018semiparametric} dataset to it. At different steps, the line is dilated to increase its thickness. Therefore, various masks are generated, see Figure \ref{fig:ablation study}. 
We apply these masks to the sample image and test the \emph{model}  against it. For each iteration, Figure \ref{fig:ablation study} illustrates the input images and results of the model.

Early steps of the model successfully recover the missing area and convert it into a plausible daytime image. After some steps, the recovered area is blurry and there is not much information to be used at the translation stage. Increasing the thickness of the line reduces the performance of both inpainting and translating as expected and the translation module is barely able to generate a daytime image. 
\section{Discussions and Future Work}

To tackle the issue of merging the image-to-image translation task and the image-inpainting task, we proposed the \emph{inpainting first} model by concatenating two existing modules and rigorously evaluated its performance after training. The model produces plausible images based on different image quality assessments both quantitatively and qualitatively. Furthermore, the ablation study indicated that the thickness input masks can drastically degrade the quality of the generated output.

An end-to-end generative network that performs both tasks simultaneously might provide more accurate results.
In this manuscript, we focused on the night to day image inpainting and provided promising results.

A deep network that is able to generate a realistic image in a modality from an image with missing points in another given modality can be used in many image enhancement tasks including medical imaging applications. 

\newpage

\section*{Acknowledgements}
This work was supported in part by the Natural Sciences and Engineering Research Council of Canada (NSERC).

\bibliographystyle{splncs04}
\bibliography{IPCV_22_3106}


\end{document}